\documentclass[12pt]{article}
\usepackage{amssymb,amsmath,amsfonts}
\usepackage{a4}
\def\b#1{\mathbf{#1}}

\def\h#1{\widehat{#1}}
\def\lra{\ \longrightarrow\ }

\def\o#1{\overline{#1}}
\def\p#1{\partial_{#1}}
\def\v#1{\vec{#1}}

\def\tr{\mathsf{Tr}}
\def\vev#1{\left\langle#1\right\rangle}
\def\D#1{$\displaystyle{#1}$}

\def\DC{\mathcal{D}}

\def\GC{\mathcal{G}}
\def\LC{\mathcal{L}}
\def\OC{\mathcal{O}}

\def\ZC{\mathcal{Z}}

\begin{document}

\title{On the definition of the covariant lattice Dirac operator}

\date{\today}

\author{Claude Roiesnel\footnote{claude.roiesnel@cpht.polytechnique.fr}}

\maketitle

\begin{center}
Centre de Physique Th\'eorique, \'Ecole Polytechnique, CNRS,\\
91128 Palaiseau cedex, France.
\end{center}

\begin{abstract}

  In the continuum the definitions of the covariant Dirac operator and
  of the gauge covariant derivative operator are tightly intertwined.
  We point out that the naive discretization of the gauge covariant
  derivative operator is related to the existence of local unitary
  operators which allow the definition of a natural lattice gauge
  covariant derivative. The associated lattice Dirac operator has all
  the properties of the classical continuum Dirac operator, in
  particular antihermiticy and chiral invariance in the massless
  limit, but is of course non-local in accordance to the
  Nielsen-Ninomiya theorem. We show that this lattice Dirac operator
  coincides in the limit of an infinite lattice volume with the naive
  gauge covariant generalization of the SLAC derivative, but contains
  non-trivial boundary terms for finite-size lattices. Its numerical
  complexity compares pretty well on finite lattices with smeared
  lattice Dirac operators.

\end{abstract}


\vfill
\begin{flushleft}
CPHT-RR 068-0812 
\end{flushleft}

\enlargethispage{0.5cm}

\newpage

\section{Introduction}

The standard mathematical description of the dynamics of the strong
interactions, a description called Quantum ChromoDynamics (QCD), is
obtained by writing the partition function of an Euclidean $SU(3)$
quantum gauge field theory interacting with $N_f\geq 2$ fermions in
the fundamental representation of the gauge group. The partition
function of QCD is postulated by analogy with the path integral
formalism of Quantum Electro-Dynamics which has been shown to be
successful with a very high accuracy. Using the rules of Grassmanian
integration, the QCD partition function can be formally written as a
functional integral over the non-abelian gauge degrees of freedom
only,
\begin{align}
  \label{qcd}
  \ZC = \int\DC A_{\mu}\left(\prod_f\det D_f\right)e^{-S_G} \,.
\end{align}
The measure of integration in eq.\,\eqref{qcd} can be interpreted as a
formal probability measure over the space of gauge configurations
because the Euclidean Dirac operator $D_f$ of each fermion flavor is
antihermitian (with the right boundary conditions) and chirally
invariant in the massless limit,
\begin{align}
  \begin{split}
    D_f = \gamma_{\mu}D_{\mu} + m_f\,,\quad
    D_{\mu} = \p{\mu} + igA_{\mu}\,,\\
    \left\{\gamma_5,\gamma_{\mu}D_{\mu}\right\} = 0\,,\quad
    \left\{\gamma_\mu,\gamma_\nu\right\} =
    2\delta_{\mu\nu}\,,\quad\gamma_\mu^\dagger=\gamma_\mu\,.
  \end{split}
\end{align}
Hence the eigenvalues of each operator $D_f$ come in complex conjugate
pairs, up to a possible discrete set of zero modes of the operator
$\gamma_{\mu}D_{\mu}$, which guarantees reality and positiveness of
their determinant for massive fermions, $\det D_f > 0$.

The Euclidean gauge action (summation over repeated indices is
implied throughout),
\begin{align}
  S_G = \frac{1}{4} F_{\mu\nu}F_{\mu\nu}\,,\quad
  F_{\mu\nu} = \p{\mu}A_{\nu}-\p{\nu}A_{\mu} +
  ig\left[A_{\mu},A_{\nu}\right]\,,  
\end{align}
is invariant under the local gauge transformations $G(x)\in SU(3)$,
\begin{align}
  \label{lgt}
  \begin{split}
    A_{\mu}(x)\lra G(x)A_{\mu}(x)G^{-1}(x) + 
    \frac{i}{g}\left(\p{\mu}G(x)\right)G^{-1}(x) \,,
  \end{split}
\end{align}
whereas the operator $D_f$ transforms covariantly.

Physical observables can then be related to expectation values with
respect to the probability measure \eqref{qcd} of certain
gauge-invariant matrix elements $\OC(A)$ of operators built out of the
Dirac operators and their inverses, provided that the formal measure
$\DC A_{\mu}$ in \eqref{qcd} be given a precise gauge invariant
meaning through a constructive procedure.

\vspace{0.5cm}

Section \ref{LR} recalls briefly the main properties of Wilson's
lattice regularization which is the only constructive proposal known
to date \cite{WIL74}. The discretization of space-time in a finite box
allows for the non-perturbative calculation of physical observables by
means of numerical simulations. The approach has been very successful
in describing many features of hadronic physics, except for one
thing. It proves difficult to reproduce the continuum physics with the
physical light quark masses.

The reason is well understood \cite{NIE81} and resides in the
formulation of lattice fermions. The Lorentz invariant regularization
of quantum fluctuations in continuum QCD generate a chiral anomaly
which cannot be duplicated on the lattice with a discretization of the
Dirac operator which is local, chirally symmetric and contains the
correct number of fermionic degrees of freedom in the continuum limit.
This result is usually referred to as the no-go theorem. The standard
avoidance is to hold to local fermions and break chiral symmetry
explicitly.

In section \ref{XD} we reconsider the naive discretization of the
covariant derivative and identify a set of local unitary operators
which enables the definition of another lattice gauge covariant
derivative with the same algebraic properties as in the
continuum. This lattice derivative is non-local in accordance to the
no-go theorem and coincides in the free limit with the SLAC derivative
\cite{DRE76}. The lattice fermion formulation based on the SLAC
derivative has been extensively discussed three decades ago. A general
consensus has emerged according to which formulations of non-local
lattice fermions coupled to gauge fields lead to various
inconsistenties in weak coupling perturbation theory
\cite{KAR78,KAR79,KAR81,RAB81,BOD87}, and cannot reproduce the
continuum limit properly, in particular the axial anomaly
\cite{NIN84}. In fact we are not aware of a single numerical study of
the functional integral of a SLAC-type fermion coupled to a compact
lattice gauge field.

Nonetheless we think it is fair to state that none of the objections
to non-local fermions has the status of a no-go theorem. In all
studies to date, the coupling of a SLAC fermion to a gauge field on
the lattice has been written by mimicking the textbook derivation of a
local gauge symmetry in the continuum. The result is correct only for
infinite lattices. The boundary conditions on finite-size lattices are
not taken into account by the standard technique. The unitary operators
exhibited in section \ref{XD} are the right tools to include boundary
conditions. As an example, in section \ref{SP} we diagonalize these
unitary operators on periodic lattices and express in section \ref{ME}
the matrix elements of the associated lattice Dirac operator in
configuration space. We find non-trivial boundary contributions which
vanish only in the limit of infinite physical volume.

The constraints of the underlying locality of the exponentiated
non-local lattice gauge covariant derivative probably cannot be
neglected in analyzing the weak coupling perturbation theory, even in
the infinite lattice volume limit. The intent of the present work is
not to address this complicated issue, which deserves a separate work,
but to stress its existence.

In the concluding remarks we point out that the numerical complexity
of the associated finite size lattice Dirac operator is similar to a
five-dimensional local Dirac operator. In fact the algorithmic
implementation is much simpler. Moreover this non-local Dirac operator
can be naturally interpreted as smeared over the Wilson lines. This
smearing has the virtue to be completely analytic. Numerical tests of
the meaningfulness of the finite-size formulation of non-local
fermions coupled to gauge fields can easily be performed on a single
desktop computer in the quenched approximation up to four dimensions.

\section{Lattice regularization}
\label{LR}

As is well-known, Wilson's formulation consists of regularizing the
Euclidean continuum gauge theory on a finite four-dimensional lattice
$\LC$ with hypercubic cells of spacing $a$, and of replacing the
continuum gauge degrees of freedom, the gauge potential $A_{\mu}(x)$
which belongs to the $SU(3)$ Lie algebra, by variables $U_{x,\mu}$
associated to each link $(x,x+a\h{\mu})$ of the lattice and which
belong to the $SU(3)$ group manifold. Then eq.\eqref{qcd} becomes
\begin{align}
  \label{lqcd}
  \ZC_\LC = \int\left(\prod_{(x,x+a\h{\mu})\in\LC}dU_{x,\mu}\right)
  \left(\prod_f\det M_f(U)\right)e^{-S_G(U)} \,.
\end{align}
where the integration measure is now a perfectly well-defined finite
product of gauge-invariant Haar measures over the $SU(3)$ group
manifold, and $S_G(U)$ and $M_f(U)$ are discretized versions of the
continuum gauge action and Dirac operators. This measure can be
evaluated numerically by stochastic importance sampling.

There is a large arbitrariness in the choice of lattice gauge action
and lattice Dirac operators. The main constraint is that the lattice
regularized model possess a second-order critical point which
reproduces the asymptotic freedom of QCD in the continuum limit, with
critical exponents predicted by perturbation theory. This requires in
particular that the lattice operators reproduce the naive continuum
definitions when the lattice spacing vanishes, $a\rightarrow
0$. Scaling theory then suggests that there exists a whole
universality class of lattice actions which correspond to different
regularizations of the same continuum theory and whose critical properties
are related by renormalization group transformations.

The most direct ab-initio approach is to consider lattice QCD actions
with the same number of parameters as continuum QCD. The simplest such
lattice gauge action, the Wilson action \cite{WIL74}, has
discretization errors of $\OC(a^2)$,
\begin{align}
  \label{wil}
  \begin{split}
    S_w(U) &= \beta\sum_{x,\mu<\nu}\left(\b{1}-\frac{1}{6}
      \tr\left(P_{\mu\nu}(x)+P^{\dagger}_{\mu\nu}(x)\right)\right)\,,
    \\
    \mathrm{with}\quad&
    P_{\mu\nu}(x) = U_{x,\mu}U_{x+a\h{\mu},\nu}
    U^{\dagger}_{x+a\h{\nu},\mu}
    U^{\dagger}_{x,\nu}\,,
    \quad\beta=\frac{6}{g^2}\,.
  \end{split}
\end{align}
Local gauge invariance is preserved on the lattice provided that the
variables $U_{x,\mu}$ transform as
\begin{align}
\label{llgt}
  U_{x,\mu}\lra G(x)U_{x,\mu}G^{-1}(x+a\h{\mu})\,.
\end{align}
The simplest candidate for a lattice Dirac operator is expressed in
terms of the naive discretization of the covariant derivative
operator, namely,
\begin{align}
  \label{asym} 
  \begin{split}
  \DC_l(U) &= \gamma_{\mu}D_{l,\mu}(U) + m\,,\\
  \left(D_{l,\mu}(U)\psi\right)_x &= \frac{1}{a}\left(U_{x,\mu}\psi_{x+a\h{\mu}} -
    \psi_x\right)\,,
  \end{split}
\end{align}
which has the correct covariant transformation law under \eqref{llgt}
but does not have a spectrum with definite transformation properties
under the conjugation operation. Hence the determinant $\det\DC_l(U)$
is complex in general and does not define a probability measure.

An obvious workaround would be to introduce antihermitian covariant
difference operators,
\begin{align}
  \label{sym}
  \left(D_{s,\mu}(U)\psi\right)_x 
  &=\frac{1}{2a}\left(U_{x,\mu}\psi_{x+a\h{\mu}} -
    U^{\dagger}_{x-a\h{\mu},\mu}\psi_{x-a\h{\mu}}\right)\,,
\end{align}
which have the same conjugation properties as the continuum operators
and produces a valid probability measure. But the operator
\D{\gamma_{\mu}D_{s,\mu}} is plagued by the famous fermion doubling
problem, due to the use of a central difference operator, and does not
describe a single fermion flavor even in the continuum limit.

Wilson proposed \cite{WIL75} to add to $D_s$ a piece proportional to
the finite difference approximation to the Laplacian operator
$\Delta$, which lifts the mass degeneracy of the fermion doublers by
terms of order $1/a$ at the expense of breaking chiral invariance
explicitly. However the Wilson operator still possess the same
pseudo-hermiticity property as the continuum Dirac operator,
\begin{align}
  \DC_w(U) = \gamma_{\mu}D_{s,\mu}(U) - r\Delta_L\quad (0<r\leq 1)\,,
  \quad \DC^{\dagger}_w = \gamma_5 \DC_w \gamma_5\,,
\end{align}
which guarantees the invariance of its spectrum under conjugacy and
the interpretation of \eqref{lqcd} as a probability measure. It was
later realized \cite{NIE81} that it is not possible to devise a
(ultra-)local lattice Dirac operator with the correct classical
continuum limit and without fermion doublers while preserving exact
chiral invariance on the lattice in the massless case.

Various alternative lattice Dirac operators have been put forward
during the subsequent three decades, some of which with a presently
viable ecosystem. The reader can find all references in a recent, and
very nice, review \cite{FOD12} of the state-of-the-art of numerical
simulations of lattice gauge theories.

\section{A chirally invariant lattice Dirac operator}
\label{XD}

If one examines the definition \eqref{asym} of the naive lattice
covariant derivative operator, one realizes immediately that the
four operators,
\begin{align}
  S_{\mu} = \b{1} + a\,D_{l,\mu}\,,\qquad\left(S_{\mu}\right)_{ix,jy} =
  \left(U_{x,x+a\h{\mu}}\right)_{ij}\delta_{y,x+a\h{\mu}}\,,
\end{align}
are unitary ladder operators which translate by one lattice unit in
direction $\h{\mu}$ each slice of the lattice field they act upon,
while rotating locally their color degrees of freedom,
\begin{align}
  S_{\mu}S_{\mu}^{\dagger} = S_{\mu}^{\dagger}S_{\mu} =
  \b{1}\,,\quad\forall\mu\,.
\end{align}
The set of operators $S_{\mu}$ transforms covariantly under the local
gauge transformations \eqref{llgt},
\begin{align}
  S_{\mu}\lra\GC S_{\mu}\GC^{-1}\,,\quad \left(\GC\right)_{ix,jy} =
  G(x)_{ij}\delta_{xy}\,,\quad\forall\mu\,,
\end{align}
and encodes all the space-time and color degrees of freedom of a gauge
field configuration on a four-dimensional lattice. For instance, the
Wilson action \eqref{wil} can be written, up to a constant term, as
\begin{align}
  S_w(U) = -\frac{\beta}{6} 
  \tr\left(S_{\mu}S_{\nu}S_{\mu}^{\dagger}S_{\nu}^{\dagger}\right)\,.
\end{align}
Expressing the unitary operators $S_{\mu}(U)$ as exponentials of
antihermitian operators $D_{r,\mu}(U)$,
\begin{align}
  S_{\mu}(U) = e^{aD_{r,\mu}(U)}\,,\quad
  D_{r,\mu}+D_{r,\mu}^{\dagger} = 0\,,\quad\forall\mu\,,
\end{align}
singles out the operators $D_{r,\mu}(U)$ as the natural definition of
the lattice covariant derivative. Indeed the symmetric covariant
difference operators \eqref{sym} are just the leading approximation in
the series expansion of these exponentials with respect to the lattice
spacing $a$,
\begin{align}
  aD_{s,\mu}(U) = \frac{1}{2}\left(e^{aD_{r,\mu}}-e^{-aD_{r,\mu}}\right)\,.
\end{align}
Then we can define the lattice Dirac operator
\begin{align}
  \DC_r(U) = \gamma_{\mu}D_{r,\mu}(U) + m\,,
\end{align}
which is antihermitian, chirally symmetric in the massless limit,
and transforms covariantly under the local gauge transformations
\eqref{llgt}. The eigenvalues of the operator $\DC_r$ come in complex
conjugate pairs, $m\pm i\lambda$, up to a possible set of zero modes
for the imaginary part which ensures, like in the continuum, reality
and positiveness of the determinant for massive fermions, $\det\DC_r >
0$.

The operator $\DC_r(U)$ is non-local since each lattice covariant
derivative operator $D_{r,\mu}$ is a series expansion in the local
covariant derivative operator $D_{l,\mu}$,
\begin{align}
\label{series}
\begin{split}
  aD_{r,\mu} &= \log(\b{1}+aD_{l,\mu}) \\
  &= \sum_{n=1}^{+\infty} \frac{(-1)^{n+1}a^n}{n}D^n_{l,\mu} \,.
\end{split}
\end{align}
From the convergence properties of the series expansion
\eqref{series}, the prospect of a practical numerical implementation
of the lattice Dirac operator $\DC_r$ might seem very slim.

On the other hand, in the free case, the eigenvectors of the operators
$S_{\mu}$ are just plane waves and their $3 N^3$ degenerate spectrum
is simply
\begin{align}
  \mathrm{Spec}\,S_{\mu}(\b{1}) = \left\{e^{iap_{\mu}}\,,\ 
    p_{\mu}=\frac{2\pi n}{aN_{\mu}}\,,\ 
    -\frac{N_{\mu}}{2}\leq n < \frac{N_{\mu}}{2}\right\} \,.
\end{align}
In this limit the lattice Dirac operator $\DC_r$ has a discrete
Fourier representation which has the same form as in the continuum,
\begin{align}
  \h{\DC}_r(\b{1}) = i\gamma_{\mu}p_{\mu} + m\,,
\end{align}
which shows that the operator $\DC_r$ is not afflicted with the
fermion doubling problem. In compliance to the Nielsen-Ninomya
theorem, the price to pay is the non-locality of the operator
$\DC_r(U)$. The operator $D_{r,\mu}(\b{1})$ coincides with the SLAC
derivative introduced long ago \cite{DRE76} which reads, in the limit
of an infinite lattice volume,
\begin{align}
  \label{slac}
  D_{\infty,\mu}(x-y) =
  \int_{-\pi/a}^{\pi/a}\frac{d^4p}{(2\pi)^4}\,ip_{\mu}\,e^{ip\cdot (x-y)} \,.
\end{align}
The usual recipe to couple a SLAC-type fermion to a $SU(3)$ gauge
field consists in restoring gauge invariance by inserting, if $x-y$
has a non-vanishing component only in direction $\h{\mu}$, the Wilson
$SU(3)$ ordered straight line integral $W_{\mu}(U,x,y)$ between $x$
and $y$,
\begin{align}
  \label{slac}
  \left(D_{S}(U)\right)_{\alpha i x,\beta j y} &= 
  m\,\delta_{\alpha \beta}\delta_{i j}\delta_{x y} +
  \sum_{\mu} \left(\gamma_{\mu}\right)_{\alpha \beta}D_{\infty,\mu}(x-y)\,
  \left(W_{\mu}(U,x,y)\right)_{i j} \,, \\
  \label{wl}
  W_{\mu}(U,x,y) &= \delta_{\b{x}^{\perp}_{\mu},\b{y}^{\perp}_{\mu}}
  \prod_{k=0}^{(y_{\mu}-x_{\mu}-1)/a}U_{x+ka\h{\mu},x+(k+1)a\h{\mu}}\,,
  \quad\mathrm{if}\ y_{\mu}>x_{\mu}\,.
\end{align}
where $x=(\b{x}^{\perp}_{\mu},x_{\mu})$,
$y=(\b{y}^{\perp}_{\mu},y_{\mu})$, and
$\b{x}^{\perp}_{\mu},\,\b{y}^{\perp}_{\mu}$ label the sites in the
three-dimensional slices orthogonal to the $\mu$th direction. If
$y_{\mu}<x_{\mu}$, we have of course
\D{W_{\mu}(U,x,y)=W_{\mu}(U,y,x)^{\dagger}}.

The matrix elements \eqref{slac} are certainly correct in the limit of
an infinite lattice. However the boundary conditions on finite-size
lattices are not taken into account by the conventional
prescription. The operators $S_{\mu}$ have such a simple structure
that it makes possible their explicit diagonalization for an arbitrary
background lattice gauge field configuration and arbitrary boundary
conditions. We shall find non-trivial boundary terms in the matrix
elements of the operator $\DC_r$ on finite-size lattices.

To the best of our knowledge, the underlying local unitary structure
of the lattice gauge covariant generalization of the SLAC derivative
does not seem to have been appreciated since its introduction.

\section{Explicit diagonalization of the unitary operators}
\label{SP}

For definiteness we shall assume the lattice to be hypercubic,
$N_1=N_2=N_3=N_4\equiv N$, and we shall impose periodic boundary
conditions on both the lattice gauge field configuration and the
lattice matter fields which can be scalars, fermions, $\cdots$. The
operators $S_{\mu}$ do not commute in general and have different
eigenspectra $\{\lambda_{\mu},\psi_{\lambda_{\mu}}\}$. Iterating the
eigenvalue equation for $S_{\mu}$,
\begin{align}
  S_{\mu}\psi_{\lambda_{\mu}} = \lambda_{\mu}\psi_{\lambda_{\mu}}\,,
\end{align}
yields
\begin{align}
  \begin{split}
  \left(S^n_{\mu}\psi_{\lambda_{\mu}}\right)_{ix} &=
  \left(\prod_{k=0}^{n-1}U_{x+ka\h{\mu},x+(k+1)a\h{\mu}}\right)_{ij}
  \left(\psi_{\lambda_{\mu}}\right)_{j,x+na\h{\mu}}\,,\\
  &= \lambda^n_{\mu}\left(\psi_{\lambda_{\mu}}\right)_{ix} \,.
\end{split}
\end{align}
Imposing the periodic boundary condition,
\begin{align}
  \psi_{x+Na\h{\mu}} = \psi_x\,,\qquad\forall x\,,
\end{align}
implies that the eigenvectors $\psi_{\lambda_{\mu}}$ satisfy the
equations
\begin{align}
  \left(W_{\mu}\left(U,x,x+Na\h{\mu}\right)\right)_{ij}
  \left(\psi_{\lambda_{\mu}}\right)_{jx} =
  \lambda^N_{\mu}\left(\psi_{\lambda_{\mu}}\right)_{ix}\,,\qquad\forall x\,,
\end{align}
where $W_{\mu}\left(U,x,x+Na\h{\mu}\right)$, defined in \eqref{wl}, is
the Wilson line from $x$ in direction $\h{\mu}$ which wraps the
lattice. We shall use the shorthand $W_{\mu,x}(U)$ for such Wilson
lines and the argument $U$ will be implicit most of the time.

Hence each non-zero space-time component of the eigenvector
$\psi_{\lambda_{\mu}}$ is a color triplet which is an eigenvector of
some Wilson line $W_{\mu,x}(U)$. The Wilson lines are covariant
objects under the local gauge transformations \eqref{llgt} and their
eigenvalues are gauge-invariant and do not depend on the choice of
base points $x$ which differ only by the $x_{\mu}$ component along
their direction. Indeed a change of base point along a Wilson line is
nothing but a similarity transformation.

Therefore the eigenspectra of the Wilson lines $W_{\mu,x}(U)$ can be
labelled by the points $\v{x}$ of the lattice slice $x_{\mu}=0$,
$\v{x}\equiv(\b{x}^{\perp}_{\mu},0)$,
\begin{align}
  \left\{e^{i\delta^c_{\mu,\v{x}}}\,,\ \eta^c_{\mu,\v{x}}\right\}\,,\quad
  -\pi < \delta^c_{\mu,\v{x}} \leq\pi\,,\quad c=1,2,3\,.
\end{align}
where $\eta^c_{\mu,\v{x}}$ are the color triplet eigenvectors of
$W_{\mu,\v{x}}(U)$. Their calculation requires only $4N^4$ $SU(3)$
matrix multiplications and $4N^3$ $SU(3)$ matrix diagonalizations.

Barring accidental degeneracies, the eigenvalues $\lambda_{\mu}$ and
eigenvectors $\psi_{\lambda_{\mu}}$ fall into families labelled by the
$4\times3\times N^3$ eigenvalues of the Wilson lines and defined by
the equations,
\begin{align}
  \lambda_{\mu}^N = e^{i\delta^c_{\mu,\v{x}}}\,.
\end{align}
The general solution for $\lambda_{\mu}$ reads
\begin{align}
  \lambda^c_{\mu,\v{x},p_{\mu}} = 
  e^{i(ap_{\mu}+\delta^c_{\mu,\v{x}}/N)}\,,\quad
  p_{\mu} = \frac{2\pi n}{Na} - \frac{\pi}{a}\,,\quad 0\leq n < N\,.
\end{align}
The $N$ non-vanishing components of the corresponding eigenvector
$\psi^c_{\mu,\v{x},p_{\mu}}$ are, with $n=0,\cdots,N-1$,
\begin{align}
\label{comp}
\begin{split}
  (\psi^c_{\mu,\v{x},p_{\mu}})_{jy} &= 
  \left(\lambda^c_{\mu,\v{x},p_{\mu}}\right)^n
  \left(\eta^c_{\mu,y}\right)_j\delta_{y,\v{x}+na\h{\mu}}\,,\\
  \mathrm{with}\qquad\eta^c_{\mu,\v{x}+y_{\mu}\h{\mu}} &= 
  W^{\dagger}_{\mu}(\v{x},\v{x}+y_{\mu}\h{\mu})\,\eta^c_{\mu,\v{x}}\,.
\end{split}
\end{align}
These components can be computed sequentially and the calculation of
each of the $12 N^4$ eigenvectors requires only $N$ $SU(3)$
matrix-vector multiplications. So the total computational complexity
of the complete diagonalization of every operator $S_{\mu}$ is of
order $\OC(N^5)$. A complete diagonalization has to be performed only
once for each lattice gauge field configuration and its storage
requirement is in practice proportional to the lattice volume since it
is more efficient to recompute the $N$ components of each eigenvector
when needed.

The eigenvectors of $S_{\mu}$ are also eigenvectors of $D_{r,\mu}$ and
the action of $D_{r,\mu}$ on an eigenvector
$\psi^c_{\mu,\v{x},p_{\mu}}$ has a remarkably simple continuum-like
expression,
\begin{align}
  \label{eigen}
  a D_{r,\mu}\psi^c_{\mu,\v{x},p_{\mu}} = i\left(ap_{\mu} + 
    \o{\alpha}^c_{\mu,\v{x}}\right)\psi^c_{\mu,\v{x},p_{\mu}}\,,
  \qquad\o{\alpha}^c_{\mu,\v{x}} = \frac{\delta^c_{\mu,\v{x}}}{N} \,.
\end{align}
There is an inherent ambiguity in the logarithmic definition of
$D_{r,\mu}$.  We have defined rather arbitrarily the phases
$\delta^c_{\mu,\v{x}}$ as the principal argument of the eigenvalues of
Wilson lines. Other prescriptions are possible for unitary gauge
groups.

\section{Matrix elements of the operator $\DC_r(U)$}
\label{ME}

The kernel operation which enters most algorithms involving fermions,
such as the calculation of the fermion propagator, is the action of
the lattice Dirac operator on an arbitrary lattice fermion field. The
action of $\DC_r$ on a fermion $\Psi$ can be written
spin-component-wise as
\begin{align}
  \left(\DC_r(U)\Psi\right)_{\alpha} =
  \left(\gamma_{\mu}\right)_{\alpha\beta}D_{r,\mu}(U)\Psi_{\beta} +
  m\Psi_{\alpha}\,.
\end{align}
To perform this calculation we just need to expand each spin component
of the fermion field over the complete eigensystem of every operator
$D_{r,\mu}$,
\begin{align}
  \label{dev}
  \Psi_{\beta} = \sum_{c,\v{x},p_{\mu}} C^c_{\beta,\mu,\v{x},p_{\mu}}
  \psi^c_{\mu,\v{x},p_{\mu}}\,,\qquad\forall\mu\,.
\end{align}
We get $4\times 4\times 3\times N^4$ equations, with
$x=(\v{x},x_{\mu})$ and $a=1$ throughout this section,
\begin{align}
  \label{sys1}
  \begin{split}
  \left(\Psi_{\beta}\right)_{jx} &= 
  \sum_{c,p_{\mu}} C^c_{\beta,\mu,\v{x},p_{\mu}}
  \left(\lambda^c_{\mu,\v{x},p_{\mu}}\right)^{x_{\mu}}
  \left(\eta^c_{\mu,x}\right)_j\,.
  \end{split}  
\end{align}
We can always choose all color triplet eigensystems
$\{\eta^c_{\mu,\v{x}}\}$ to be orthonormal. Then we
observe that all eigensystems $\{\eta^c_{\mu,x}\}$ along the same
Wilson line are simultaneously orthonormal,
\begin{align}
  \sum_j\left(\eta^{\star a}_{\mu,x}\right)_j\left(\eta^b_{\mu,x}\right)_j =
  \delta^{ab}\,,\quad\forall x=(\v{x},x_{\mu})\,,
\end{align}
since they are related by unitary transformations which preserve the
scalar product. Thus we can transform each equation \eqref{sys1} into
a simple one-dimensional Fourier series,
\begin{align}
  \begin{split}
  \left(\Psi'^c_{\beta}\right)_x &= e^{-ix_{\mu}\o{\alpha}^c_{\mu,\v{x}}}
  \sum_j\left(\eta^{\star c}_{\mu,x}\right)_j\left(\Psi_{\beta}\right)_{jx}\,,\\
  &= \sum_{p_{\mu}} C^c_{\beta,\mu,\v{x},p_{\mu}} e^{ip_{\mu}x_{\mu}}\,,
  \quad\forall x=(\v{x},x_{\mu}) \,.
\end{split}
\end{align}
Therefore the coefficients $C_{\beta,c,\mu,\v{x},p_{\mu}}$ are the
one-dimensional inverse discrete Fourier transforms,
\begin{align}
  \label{inv}
  \begin{split}
    C^c_{\beta,\mu,\v{x},p_{\mu}} &= \frac{1}{N}
    \sum_{x_{\mu}}\left(\Psi'^c_{\beta}\right)_x e^{-ip_{\mu}x_{\mu}}\,.
  \end{split}
\end{align}
It follows that the total computational complexity of the action of
the operator $\DC_r(U)$ on a fermion field is of order $\OC(N^5)$
which is only a factor $N$ more expensive than the action of a a local
operator like the Wilson operator $\DC_w(U)$.

Plugging \eqref{inv} into \eqref{dev} yields, with $x=\v{x}+x_{\mu}\h{\mu}$,
\begin{align}
  \left(D_{r,\mu}\Psi_{\beta}\right)_{jy} = \frac{1}{N}\sum_{c,\v{x},x_{\mu},p_{\mu}}
  i(p_{\mu}+\o{\alpha}^c_{\mu,\v{x}})e^{-i(p_{\mu}+\o{\alpha}^c_{\mu,\v{x}})x_{\mu}}
  \left(\sum_k\left(\eta^{\star c}_{\mu,x}\right)_k
    \left(\Psi_{\beta}\right)_{kx}\right)\left(\psi^c_{\mu,\v{x},p_{\mu}}\right)_{jy}
\end{align}
Inserting \eqref{comp} gives, with $y=\v{y}+y_{\mu}\h{\mu}$,
\begin{align}
  \left(D_{r,\mu}\Psi_{\beta}\right)_{jy} =
  \frac{i}{N}
  \sum_{c,\v{x},x_{\mu},p_{\mu}}\delta_{\v{y},\v{x}}\,
  (p_{\mu}+\o{\alpha}^c_{\mu,\v{x}})
  e^{i(p_{\mu}+\o{\alpha}^c_{\mu,\v{x}})(y_{\mu}-x_{\mu})}
  \left(\sum_k\left(\eta^{\star c}_{\mu,x}\right)_k
    \left(\Psi_{\beta}\right)_{kx}\right)\left(\eta^c_{\mu,y}\right)_j
\end{align}
The summation over $p_{\mu}$ brings in the finite-size SLAC
derivative,
\begin{align}
  \label{slacN}
  D_{N,\mu}(x_{\mu}) = \frac{1}{N}\sum_{p_{\mu}} 
  ip_{\mu}\,e^{ip_{\mu}x_{\mu}}\,,
\end{align}
and the summation over $\v{x}$ produces,
\begin{align}
\begin{split}
  \left(D_{r,\mu}\Psi_{\beta}\right)_{jy} &= \delta_{\v{x},\v{y}}
  \sum_{x_{\mu}} D_{N,\mu}(y_{\mu}-x_{\mu})
  \sum_k\left(\sum_ce^{i\o{\alpha}^c_{\mu,\v{y}}(y_{\mu}-x_{\mu})}
    \left(\eta^{\star c}_{\mu,x}\right)_k\left(\eta^c_{\mu,y}\right)_j
    \right)\left(\Psi_{\beta}\right)_{kx}  \\
  &+ i\sum_k\left(\sum_c\o{\alpha}^c_{\mu,\v{y}}
    \left(\eta^{\star c}_{\mu,y}\right)_k
    \left(\eta^c_{\mu,y}\right)_j\right) \left(\Psi_{\beta}\right)_{ky}
\end{split}
\end{align}
The last step is to insert \eqref{comp} and use the identity,
\begin{align}
  \left(W^{\dagger}_{\mu}\left(\v{y},\v{y}+y_{\mu}\h{\mu}\right)\right)_{jl}
  \left(\sum_{c}f\left(e^{i\delta^c_{\mu,\v{y}}}\right)
    \left(\eta^c_{\mu,\v{y}}\right)_l
    \left(\eta^{\star c}_{\mu,\v{y}}\right)_m\right)
  \left(W_{\mu}\left(\v{y},\v{y}+y_{\mu}\h{\mu}\right)\right)_{mk}
   =  \left(f\left(W_{\mu,y}\right)\right)_{jk} \,.
\end{align}
Collecting everything together, the matrix elements of the operator
$\DC_r(U)$ read finally,
\begin{align}
  \label{Dr}
  \begin{split}
  \left(\DC_r(U)\right)_{\alpha j y,\beta i x} &= 
  \delta_{x y}\left(m\,\delta_{\alpha \beta}\delta_{i j} + \frac{1}{Na}
    \sum_{\mu}\left(\gamma_{\mu}\right)_{\alpha \beta} 
    \left(\log\left(W_{\mu,y}(U)\right)\right)_{ji}\right) + 
  \delta_{\v{x},\v{y}} \times \\
  &\times \sum_{\mu}
  \left(\gamma_{\mu}\right)_{\alpha \beta}D_{N,\mu}(y_{\mu}-x_{\mu})\,
  \left(W_{\mu}(U,y,x)
    \left(W_{\mu,x}(U)\right)^{(y_{\mu}-x_{\mu})/Na}\right)_{j i} \,,
  \end{split}
\end{align}
where we reintroduced the lattice spacing $a$ to make apparent the
physical dimensions. The covariance of the operator $\DC_r(U)$ under
the local gauge transformations \eqref{llgt} is clearly satisfied,
\begin{align}
  \DC_r(U)\quad\lra\quad\GC\,\DC_r(U)\,\GC^{-1}\,.
\end{align}
We find two additional boundary contributions with respect to the
infinite volume expression \eqref{slac}. The first one is a diagonal
term in configuration space which vanishes proportionally to the
inverse physical lattice size. The second one is an insertion in the
open Wilson line $W_{\mu}(U,y,x)$ of the closed Wilson line
$W_{\mu,x}(U)$ raised to a power the variation of which is also
proportional to the distance $y-x$ in physical units. The insertion
point can be covariantly transported anywhere along the closed Wilson
line.

\section{Outlook}

As recalled in the introductory section, the lattice fermion
formulation based on the SLAC derivative has been rather
controversial. There have been one-loop calculations with the SLAC
operator \eqref{slac} in weak coupling perturbation theory of lattice
Quantum ElectroDynamics (QED) in four dimensions, which have shown the
occurence of singularities in the fermion triangle graph \cite{KAR78}
and in the vacuum polarization \cite{KAR79}, that lead to non-local,
non Lorentz covariant expressions. These singularities are generated
by the discontinuities of the SLAC derivative at the edges of the
Brillouin zone, $p_{\mu}=\pm \pi/a$. It has been claimed \cite{KAR81}
that these divergences could not be renormalized while keeping in the
continuum limit $a\rightarrow 0$ (at $L=Na\rightarrow\infty)$,
both chiral invariance without extra states and Lorentz
invariance. However it has also been suggested \cite{RAB81} that QED
could be recovered in the continuum limit by a proper,
non-perturbative, treatment of the infrared singularities and by
imposing a finite number of non-local renormalization conditions.  But
the application of such empirical prescriptions to the lattice
Schwinger model, namely two-dimensional Quantum Electrodynamics with
massless fermions, which is a completely solvable model in the
continuum \cite{SCH62}, has still generated a spectrum doubling, a
vanishing anomaly, a vanishing vacuum expectation value for
$\vev{\o{\psi}\psi}$, and a non-covariant axial-vector \cite{BOD87}.

Despite all these negative results, a rigorous treatment of the
perturbative expansion of the lattice Dirac operator \eqref{slac} in
the infinite volume limit is still lacking. As already emphasized, the
underlying unitarity structure of the non-local lattice covariant
derivative has not been taken into account in existing studies and the
issue of spectrum doubling in such a formulation, which depends
crucially upon the handling of singularities in the infinite volume
limit, has to be settled accordingly.

Whatever the outcome, we advocate a pragmatic approach, \`a-la Wilson.
We have exhibited a covariant and chirally-invariant lattice Dirac
operator on finite-size lattices which has certainly no spectrum
doubling in the free limit. Moreover, the expressions \eqref{Dr} of
the matrix elements of the operator $\DC_r$ are quite convenient for
an actual computer implementation. If spectrum doubling does occur
when the gauge interaction is turned on, one could always add an
explicit chiral symmetry breaking term. Our derivation reveals that
the non-local operator $\DC_r$ is a smeared operator with a
controllable analytic averaging over the links of Wilson lines
(whereas the original smearing proposal \cite{APE87}, as well as its
many variants, are empirical thickenings of the links). Smeared
operators are smoother and it is widely known \cite{FOD12} that their
inversion has better convergence properties than local operators. So
the accelerated convergence near the chiral limit, whose qualitative
nature can be studied in the quenched approximation, may even turn out
to compensate for the additional computational complexity of $\DC_r$
with respect to local Dirac operators such as $\DC_w$.

\end{document}